\documentstyle[aas2pp4]{article}

\def\Ha{H$\alpha$}

\def\etal{et\thinspace al.~}
\def\eg{e.g.,~}

\def\msol{\rm\,M_\odot}
\def\Rsol{\rm\,R_\odot}
\def\Lsol{\rm\,L_\odot}
\def\ergs{{\rm\,erg\,s^{-1}}}
\def\kms{{\rm\,km\,s^{-1}}}

\def\spose#1{\hbox to 0pt{#1\hss}}
\def\Dt{\spose{\raise 1.5ex\hbox{\hskip3pt$\mathchar"201$}}}    

\def\Mdot{\Dt M}

\received{11 March 1998}
\accepted{28 May 1998}

\lefthead{M. S. Oey and S. A. Smedley}
\righthead{Shell and Star Formation}

\begin{document}

\title{ Shell formation and star formation in superbubble DEM 192 \\
\medskip\small Accepted 28-05-98 to the Astronomical Journal}

\author{M. S. Oey\altaffilmark{1} }
\affil{Institute of Astronomy, Madingley Road, Cambridge, CB3 0HA,
	U.K.; oey@ast.cam.ac.uk }
\and

\author{Shona A. Smedley\altaffilmark{2}}
\affil{Dept. of Mathematics and Computer Science, University of Leicester,
	University Road, Leicester, LE1 7RH, U.K.; sas10@le.ac.uk}

\altaffiltext{1}{Visiting Astronomer, Cerro Tololo Inter-American
	Observatory, National Optical Astronomy Observatories, operated by the
	Association of Universities for Research in Astronomy, Inc.,
	under a cooperative agreement with the National Science Foundation.}
\altaffiltext{2}{Participant in the 1997 Summer Student Course of the Royal
	Greenwich Observatory.}


\begin{abstract}
Was star formation in the OB associations, LH~51 and LH~54, triggered
by the growth of the superbubble DEM~192?  To examine this possibility,
we investigate the stellar contents and star formation history, and
model the evolution of the shell.  H-R diagrams constructed from
$UBV$ photometry and spectral classifications indicate highly coeval
star formation, with the entire
massive star population having an age of $\lesssim$ 2--3 Myr.  However,
LH~54 is constrained to an age of $\sim$ 3 Myr by the presence of a WR
star, and the IMF for LH~51 suggests a lower-mass limit implying
an age of 1--2 Myr.  There is no evidence of an earlier stellar
population to create the superbubble, but the modeled shell kinematics
are consistent with an origin due to the strongest stellar winds of LH~54.
It might therefore be possible that LH~54 created the superbubble, which in
turn may have triggered the creation of LH~51.  Within the errors, the
spatial distribution of stellar masses and IMF appear uniform within
the associations.

We reinvestigate
the estimates for stellar wind power $L_w(t)$, during the H-burning
phase, and note that revised mass-loss rates yield a significantly
different form for $L_w(t)$, and may affect stellar evolution
timescales.  We also model superbubble expansion into an ambient
medium with a sudden, discontinuous drop in density, and find that
this can easily reproduce the anomalously high shell expansion
velocities seen in many superbubbles. 
\end{abstract}



\section{Introduction}

As early as the work of Bok, Bok, \& Basinski (1962), the OB
associations LH 51 and LH 54 (Lucke \& Hodge 1970) in the Large
Magellanic Cloud (LMC) have been
recognized as an extremely young star-forming region.  However, the
complex is especially notable for its associated superbubble, which is
one of the most prominent such objects in the LMC that is seen in \Ha\ 
(\eg Meaburn 1980).  This nebula has been
cataloged in \Ha\ surveys of the LMC as DEM 192 (Davies, Elliott, \&
Meaburn 1976) and N51 D (Henize 1956).   Figure~1 shows
images kindly obtained by R.C. Smith with the U. Michigan/CTIO Curtis
Schmidt Telescope, in the light of \Ha\ (Figure~1$a$, Plate *) and
[\ion{S}{2}] (Figure~1$b$, Plate *).  Interestingly, the 
young, massive stars are not centered within the the superbubble, but
are located toward the east and west edges of the nebular shell.  The
geometry of the complex is therefore suggestive of triggered, or
propagating star formation induced by the accumulation and compression
of gas in the expanding shell.

\begin{figure*}
\figurenum{1$a$}
\begin{center}
\epsfbox[60 160 500 570]{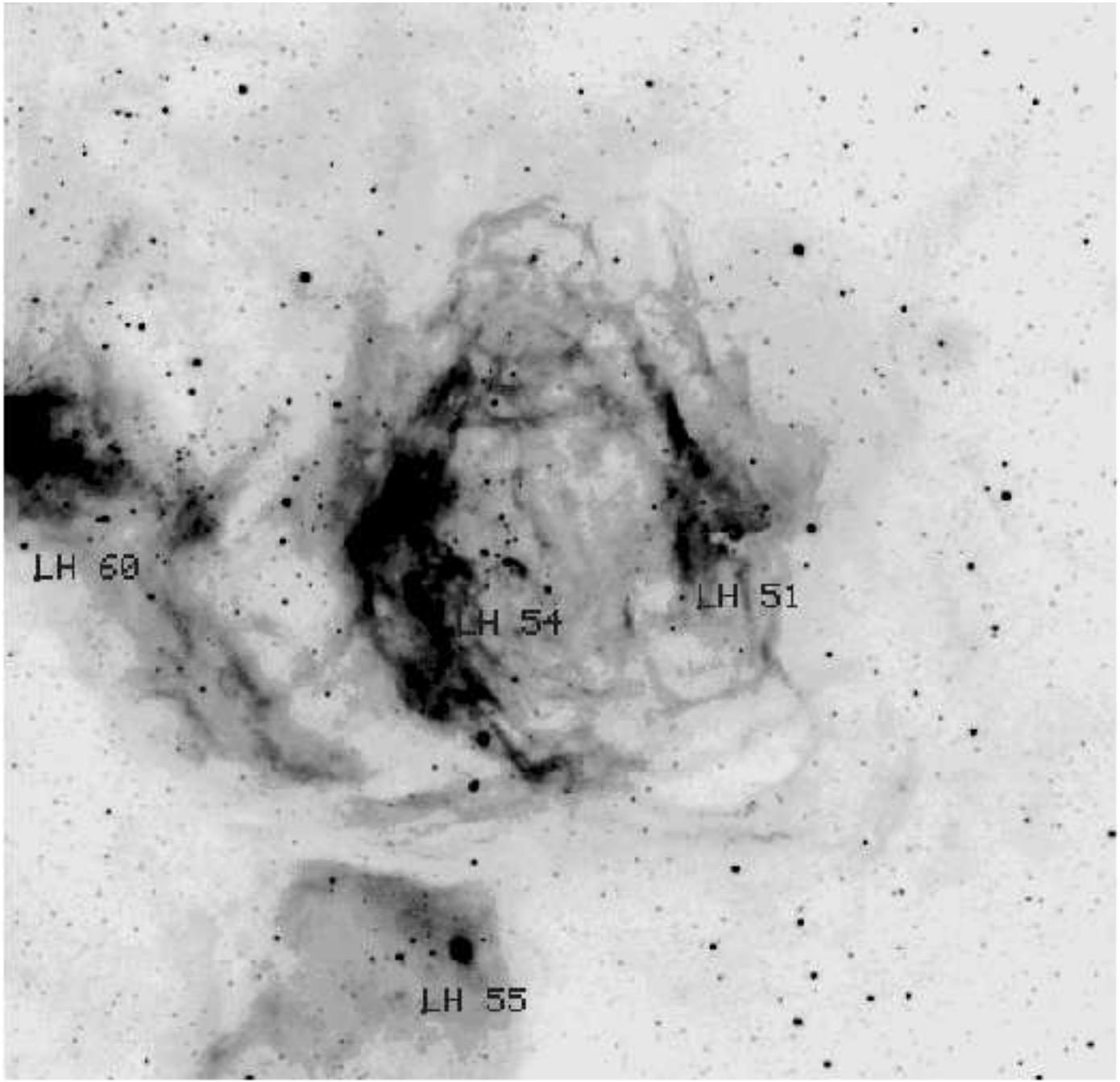}
\end{center}
\caption{DEM 192 in \Ha, provided by R. C. Smith from the
U. Michigan/CTIO Magellanic Clouds Emission-Line Survey.
\label{Ha}}
\end{figure*}

\begin{figure*}
\figurenum{1$b$}
\begin{center}
\epsfbox[60 160 500 570]{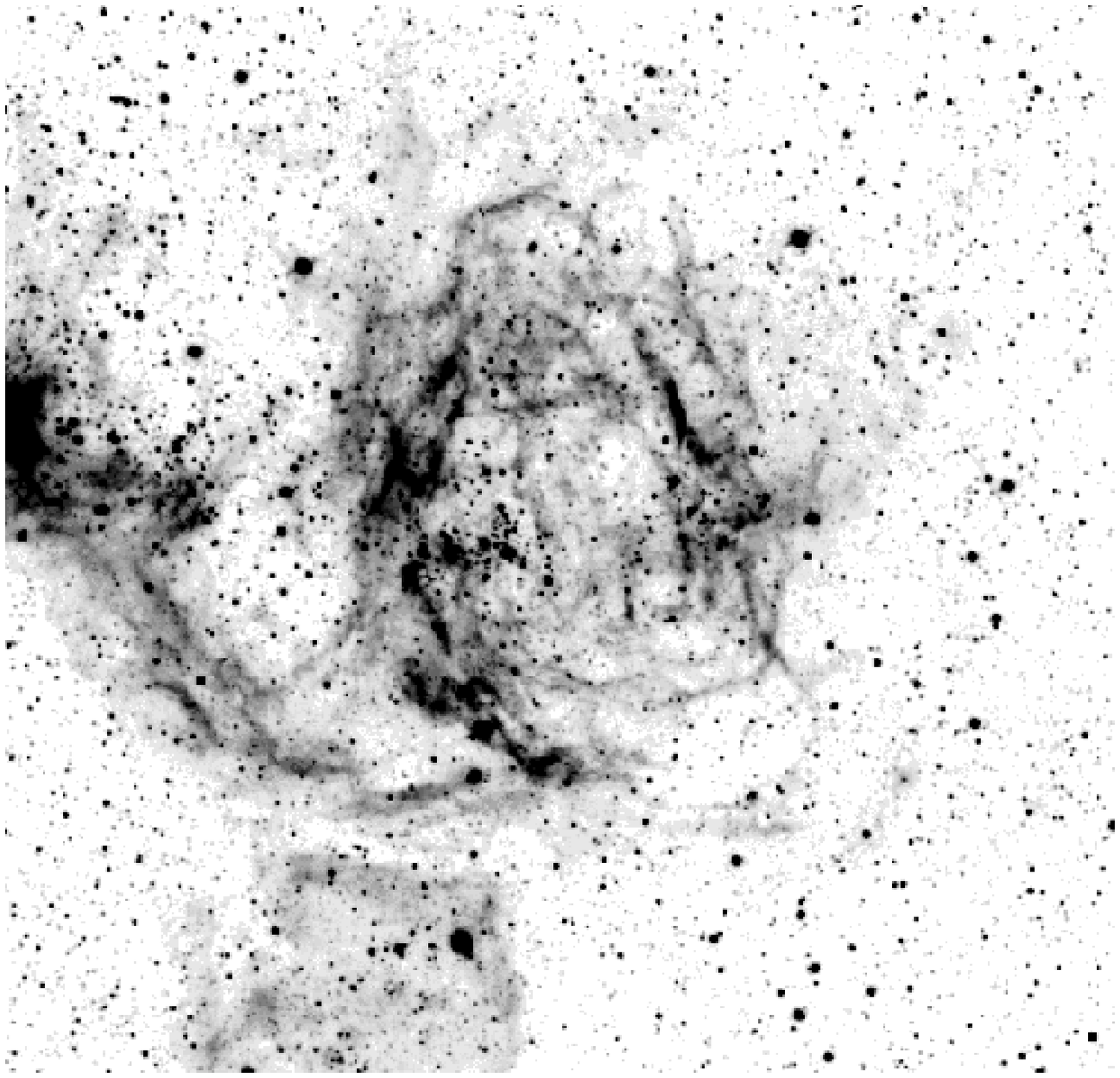}
\end{center}
\caption{DEM 192 in [\ion{S}{2}], also obtained by R. C. Smith. }
\end{figure*}
\setcounter{figure}{1}

Since both OB associations, especially LH 54, are aligned radially
with respect to the shell, it is particularly suggestive that star
formation may be propagating outward as the superbubble expands.  If
this were the case, then quantifying the age gradient would be of
primary interest in understanding the progression of triggered star
formation, as well as confirming superbubble expansion timescales,
which are not well-understood (Oey 1996a).  We therefore wish
to determine the star formation history in this complex, and shell
evolution, in relation to each other.  This will require a
detailed investigation of the stellar population, with subsequent
modeling of the superbubble.  The stellar data will also allow us 
to examine the initial mass function (IMF) across the region.

\section{Observations}

We have presented $UBV$ photometry for 1460 stars in this region in a
previous study (Oey 1996b), where a CCD mosaic image\footnote{The CCD
mosaic image is available on the AAS CDROM series, Vol. 6} of the
stellar population may be found, which identifies the visually
brightest members.  Subsequent spectroscopic observations of the hottest stars 
were obtained as part of the sample studied by Oey (1996c).  The
observational details may be found in that paper, so we summarize them
only briefly here.  

\begin{figure*}
\begin{center}
\epsfbox[60 280 500 520]{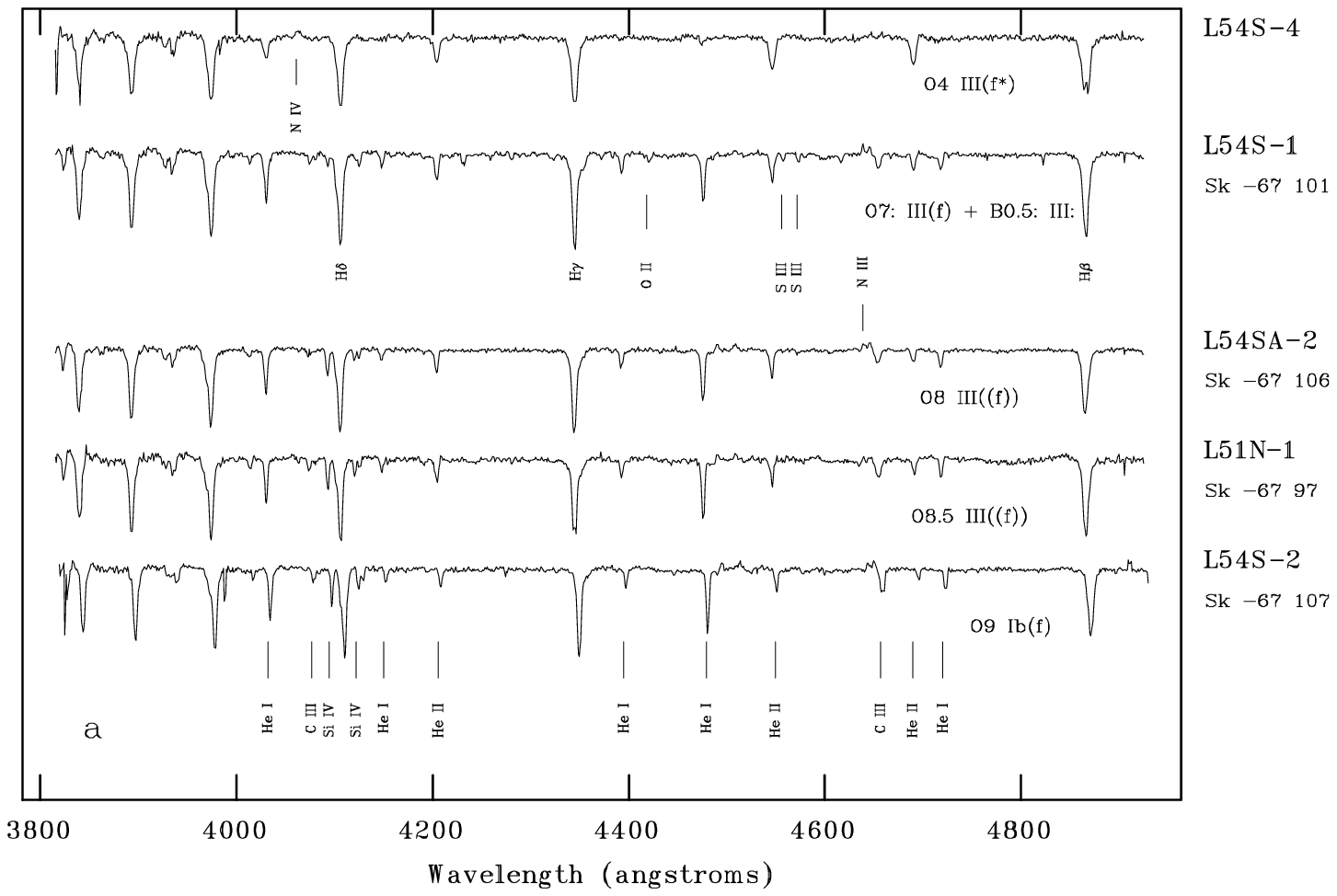}
\end{center}
\end{figure*}

\begin{figure*}
\begin{center}
\epsfbox[60 280 500 530]{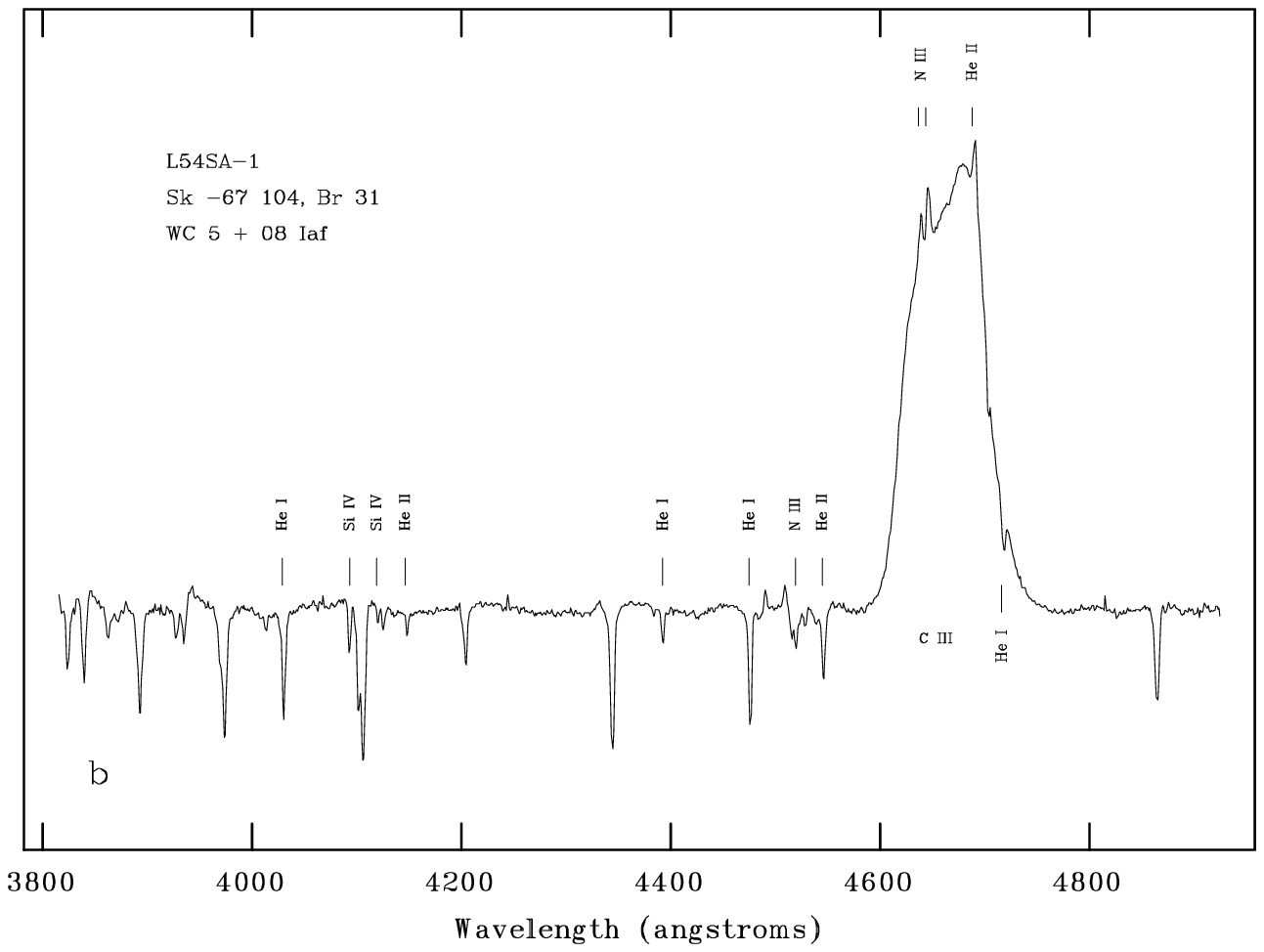}
\end{center}
\caption{Classification spectra of the visually brightest stars.
\label{spectra}}
\end{figure*}

The spectra were obtained in 1994 January at the
CTIO 4-m Blanco telescope, using the Argus multi-fiber spectrograph.
A 632 line mm$^{-1}$ grating was used in second order, yielding a
wavelength coverage of about 3800--4900 \AA.  The spectra, having a
resolution of $2.2$ \AA, were recorded on a Reticon II 1200 $\times$
400 pixel CCD.  After standard CCD reduction procedures, the target
observations in each exposure were sky-subtracted with spectra
combined from the 24 or more fibers set on sky positions.  The target
spectra were then rectified and classified according to the criteria
of Walborn \& Fitzpatrick (1990).  A few spectra of the brightest
stars are shown in Figure~\ref{spectra}$a$ and $b$.

In keeping with the companion study of OB associations, the stellar
data were transformed to $M_{\rm bol}$ and $T_{\rm eff}$ according to
the prescription in that work (Oey 1996c).  For stars with spectral
types, we used the $T_{\rm eff}$ calibrations and bolometric
corrections of Chlebowski \& Garmany (1991) and Humphreys \& McElroy
(1984).  Stars with only photometric data were transformed with
the relations given by Massey {\etal}(1995).  As described in Oey
(1996c), stars with errors in color $> 0.07$ mag have $T_{\rm eff}$
derived from their $B-V$ color alone, while those with more reliable
colors use both $B-V$ and $U-B$ to obtain $T_{\rm eff}$.  We find a
median and mean reddening of $E(B-V)= 0.08$ and 0.09, respectively,
derived from the 141 
stars with combined photometric errors $\lesssim 0.05$ mag.  Since the
reddening distribution is asymmetric, we adopt the median value for
all stars for which $T_{\rm eff}$ was derived
from the $B-V$ color only.  We use the extinction function of Savage
\& Mathis (1979), with $A_V = 3.1\times E(B-V)$ and $E(U-B) = 0.66
\times E(B-V)$.  As in the other papers in this series, we
adopted a distance modulus to the LMC of 18.4.

Table~\ref{data} presents empirical
parameters for the 47 stars in the spectroscopic sample.  The stars are
identified in the first column, according to the nomenclature of Oey
(1996b), where photometric data for the entire sample may be found.  
In columns~2 and 3, we list the stellar coordinates for epoch J2000.0.
Columns 4 -- 6 give the photometric data, uncorrected for extinction:
$V$, $B-V$ and $U-B$.
Columns 7 -- 10 give the color excess $E(B-V)$, $\log\ T_{\rm eff}$,
$M_{\rm bol}$, and the spectral classification.
Three spectroscopic binaries were treated as described in Oey (1996c),
and are identified in Table~\ref{data} with ``a'' and ``b'' notations.
Among these is the WC~5 star Br~31 \break (Sk --67~104), whose $T_{\rm eff}$
and $M_{\rm bol}$ are indeterminate.  Its companion had been
classified as O9 by Breysacher (1981), and we refine this to O8~Iaf
(Figure~\ref{spectra}$b$).  Three additional binary candidates are
noted with the ``\#'' symbol on their spectral classifications.
A few bright stars with identifications from other catalogs are
cross-referenced in Oey (1996b).

\begin{deluxetable}{lcccccccl}
\scriptsize
\tablecaption{Stellar Parameters of Stars with Spectral Types \label{data}}
\tablewidth{0pt}
\tablehead{
\colhead{Star ID} & \colhead{R.A.~~(J2000)~~Dec.} & 
\colhead{$V$} & \colhead{$B-V$} & 
\colhead{$U-B$} & \colhead{$E(B-V)$} & 
\colhead{$\log T_{\rm eff}$}  & \colhead{$M_{\rm bol}$} & 
\colhead{Sp. Type\tablenotemark{a}}
} 
\startdata
L54SA-1b   & 5~26~04.31 ~ --67~29~58.7 & 11.418 & --0.118 & --1.014 & 0.08 & 4.53 & --10.5\phn & O8 Iaf     \\
L54SA-2    & 5~26~15.51 ~ --67~30~01.8 & 11.963 & --0.158 & --1.048 & 0.15 & 4.56 & --10.4\phn & O8 III((f)) \\
L54S-2     & 5~26~20.91 ~ --67~29~57.7 & 12.674 & --0.165 & --1.021 & 0.12 & 4.51 &  --9.2 & O9 Ib(f)   \\
L51N-1     & 5~25~15.84 ~ --67~28~06.5 & 13.019 & --0.199 & --1.006 & 0.11 & 4.54 &  --9.1 & O8.5 III((f)) \\
L54S-1a    & 5~25~56.57 ~ --67~30~30.1 & 13.111 & --0.187 & --1.021 & 0.08 & 4.58 &  --9.2 & O7: III(f) \\
L54S-4     & 5~26~24.27 ~ --67~30~17.4 & 13.130 & --0.215 & --1.010 & 0.10 & 4.65 &  --9.6 & O4 III(f*) \\
L54S-5     & 5~26~15.80 ~ --67~29~44.3 & 13.159 & --0.173 & --1.054 & 0.14 & 4.56 &  --9.1 & O8 III((f)) \\
L54S-3a    & 5~26~08.95 ~ --67~29~47.5 & 13.311 & --0.218 & --1.008 & 0.08 & 4.61 &  --9.2 & O6.5: V    \\
L54S-3b    & 5~26~08.95 ~ --67~29~47.5 & 13.511 & --0.218 & --1.008 & 0.08 & 4.26 &  --6.7 & B2: II:    \\
L54S-1b    & 5~25~56.57 ~ --67~30~30.1 & 13.611 & --0.187 & --1.021 & 0.08 & 4.41 &  --7.5 & B0.5: III: \\
L54N-2     & 5~26~08.19 ~ --67~28~28.3 & 13.739 & --0.190 & --0.979 & 0.13 & 4.61 &  --8.9 & O6.5 V     \\
L54N-3     & 5~26~23.75 ~ --67~29~07.1 & 13.841 & --0.139 & --0.928 & 0.12 & 4.32 &  --6.9 & B1 IIIe    \\
L54N-4     & 5~26~05.47 ~ --67~29~07.7 & 13.909 & --0.196 & --1.037 & 0.11 & 4.56 &  --8.3 & O9 V       \\
L54S-8     & 5~26~32.16 ~ --67~29~51.3 & 13.966 & --0.178 & --0.872 & 0.04 & 4.36 &  --6.6 & B0.5 Ib    \\
L51N-3     & 5~25~31.03 ~ --67~28~38.9 & 13.992 & --0.223 & --0.999 & 0.09 & 4.56 &  --8.1 & O9 V       \\
L54S-9     & 5~26~20.78 ~ --67~29~52.1 & 14.066 & --0.197 & --0.882 & 0.10 & 4.48 &  --7.6 & B0 III     \\
L54N-6     & 5~26~28.13 ~ --67~27~23.1 & 14.310 & --0.189 & --0.992 & 0.12 & 4.56 &  --7.9 & O9 V       \\
L54S-10    & 5~26~03.31 ~ --67~32~10.4 & 14.355 & {\phn}0.045 & --1.097 & 0.30 & 4.38 &  --7.5 & early B:: +neb\\
L54N-7     & 5~26~18.51 ~ --67~26~45.6 & 14.369 & --0.063 & --0.963 & 0.22 & 4.40 &  --7.3 & B0.5:: V +neb \\
L51S-2     & 5~25~28.42 ~ --67~29~13.3 & 14.418 & --0.221 & --1.043 & 0.10 & 4.63 &  --8.2 & O6 V((f))e \\
L54SA-4    & 5~26~15.85 ~ --67~29~49.9 & 14.472 & --0.217 & --0.941 & 0.06 & 4.41 &  --6.6 & B0.5 III   \\
L54S-12    & 5~26~04.42 ~ --67~29~35.6 & 14.624 & --0.196 & --0.962 & 0.10 & 4.47 &  --7.1 & B0 V \#     \\
L54S-13    & 5~26~19.98 ~ --67~30~01.7 & 14.640 & --0.191 & --0.896 & 0.09 & 4.40 &  --6.6 & B0.5 Ve    \\
L54S-14    & 5~26~16.85 ~ --67~29~38.0 & 14.641 & --0.197 & --0.976 & 0.10 & 4.54 &  --7.5 & O9.5 V     \\
L54S-15    & 5~26~24.13 ~ --67~30~33.6 & 14.645 & --0.222 & --1.007 & 0.09 & 4.58 &  --7.6 & O8 Ve      \\
L54S-16    & 5~26~02.91 ~ --67~29~33.2 & 14.646 & --0.237 & --0.979 & 0.06 & 4.54 &  --7.3 & O9.5 V \#  \\
L51S-3     & 5~25~27.36 ~ --67~29~10.0 & 14.675 & --0.239 & --1.019 & 0.07 & 4.58 &  --7.6 & O8 V       \\
L51S-4     & 5~25~35.31 ~ --67~28~56.7 & 14.688 & --0.222 & --0.960 & 0.06 & 4.40 &  --6.5 & B0.5 V     \\
L54S-18    & 5~26~14.43 ~ --67~29~34.6 & 14.692 & --0.229 & --0.998 & 0.08 & 4.56 &  --7.4 & O9 V       \\
L54S-19    & 5~26~06.45 ~ --67~29~50.7 & 14.820 & --0.114 & --1.004 & 0.17 & 4.40 &  --6.7 & B0.5: V \#  \\
L51S-5     & 5~25~30.74 ~ --67~29~26.1 & 14.948 & --0.225 & --0.953 & 0.06 & 4.40 &  --6.2 & B0.5 V     \\
L51N-9     & 5~25~31.86 ~ --67~26~51.7 & 14.999 & --0.219 & --0.909 & 0.04 & 4.38 &  --6.0 & B1: V +neb \\
L51N-10    & 5~25~32.50 ~ --67~28~37.5 & 15.041 & --0.211 & --0.967 & 0.09 & 4.47 &  --6.6 & B0 V       \\
L54S-20    & 5~26~08.51 ~ --67~30~25.6 & 15.078 & --0.199 & --1.010 & 0.10 & 4.54 &  --7.0 & O9.5 V     \\
L54SA-5    & 5~26~15.14 ~ --67~30~06.2 & 15.087 & --0.210 & --0.815 & 0.05 & 4.38 &  --6.0 & B1:: V +neb \\
L51S-7     & 5~25~14.19 ~ --67~29~07.0 & 15.101 & --0.203 & --0.946 & 0.10 & 4.54 &  --7.0 & O9.5 V     \\
L54SA-1a   & 5~26~04.31 ~ --67~29~58.7 & 15.118 & --0.118 & --1.014 & 0.08 & \nodata & \nodata & WC 5     \\
L51N-12    & 5~25~28.17 ~ --67~27~59.0 & 15.121 & --0.218 & --0.943 & 0.06 & 4.40 &  --6.1 & B0.5 V +neb \\
L54S-22    & 5~26~04.20 ~ --67~29~25.5 & 15.188 & --0.156 & --0.972 & 0.14 & 4.54 &  --7.0 & O9.5-B0 V  \\
L54S-24    & 5~26~21.81 ~ --67~29~24.2 & 15.264 & --0.211 & --0.993 & 0.09 & 4.54 &  --6.8 & O9.5 V +neb \\
L54N-14    & 5~25~56.15 ~ --67~28~10.5 & 15.279 & --0.204 & --0.990 & 0.10 & 4.54 &  --6.8 & O9.5 V     \\
L54S-25    & 5~26~17.93 ~ --67~30~42.5 & 15.288 & --0.190 & --0.885 & 0.06 & 4.30 &  --5.1 & B1.5 III +neb \\
L54N-16    & 5~26~14.25 ~ --67~28~43.6 & 15.313 & --0.196 & --0.975 & 0.10 & 4.54 &  --6.8 & O9.5 V     \\
L54S-26    & 5~26~12.42 ~ --67~30~45.1 & 15.408 & --0.197 & --0.991 & 0.06 & 4.38 &  --5.7 & B1:: V +neb \\
L54N-18    & 5~26~08.80 ~ --67~26~34.8 & 15.461 & {\phn}0.052 & --0.903 & 0.37 & 4.63 &  --8.0 & O6: Ve     \\
L51S-9     & 5~25~40.62 ~ --67~30~40.9 & 15.473 & --0.238 & --0.919 & 0.06 & 4.47 &  --6.1 & B0:: V     \\
L54SA-7    & 5~26~05.06 ~ --67~29~58.5 & 15.499 & --0.260 & --0.735 & 0.03 & 4.34 &  --5.2 & B1.5 V     \\
L54N-19    & 5~26~14.30 ~ --67~27~11.2 & 15.527 & --0.169 & --0.993 & 0.13 & 4.54 &  --6.7 & O9.5 V     \\
L54N-21    & 5~25~53.87 ~ --67~26~43.4 & 15.593 & --0.152 & --0.797 & 0.07 & 4.28 &  --4.9 & B2.5 V     \\
L54S-36    & 5~26~09.22 ~ --67~31~11.1 & 15.789 & --0.184 & --0.967 & 0.07 & 4.34 &  --5.0 & B1.5:: Ve  \\
L54N-25    & 5~26~24.95 ~ --67~28~53.0 & 15.816 & --0.002 & --0.854 & 0.24 & 4.29 &  --5.3 & B2:: V +neb \\
L54N-27    & 5~26~29.30 ~ --67~27~09.9 & 15.921 & --0.149 & --0.946 & 0.15 & 4.47 &  --5.9 & O9.5-early Be\\
L51S-14    & 5~25~21.44 ~ --67~31~26.6 & 15.960 & --0.201 & --0.882 & 0.06 & 4.38 &  --5.1 & B1:: V     \\
L54N-33    & 5~25~55.86 ~ --67~27~44.4 & 16.390 & --0.164 & --0.939 & 0.10 & 4.38 &  --4.8 & B1:: V     \\
\enddata
\tablenotetext{a}{The \# symbol indicates candidates for spectroscopic binaries.}

\end{deluxetable}

\section{The Stellar Population and Formation History}

\begin{figure*}
\begin{center}
\epsfbox[20 430 500 600]{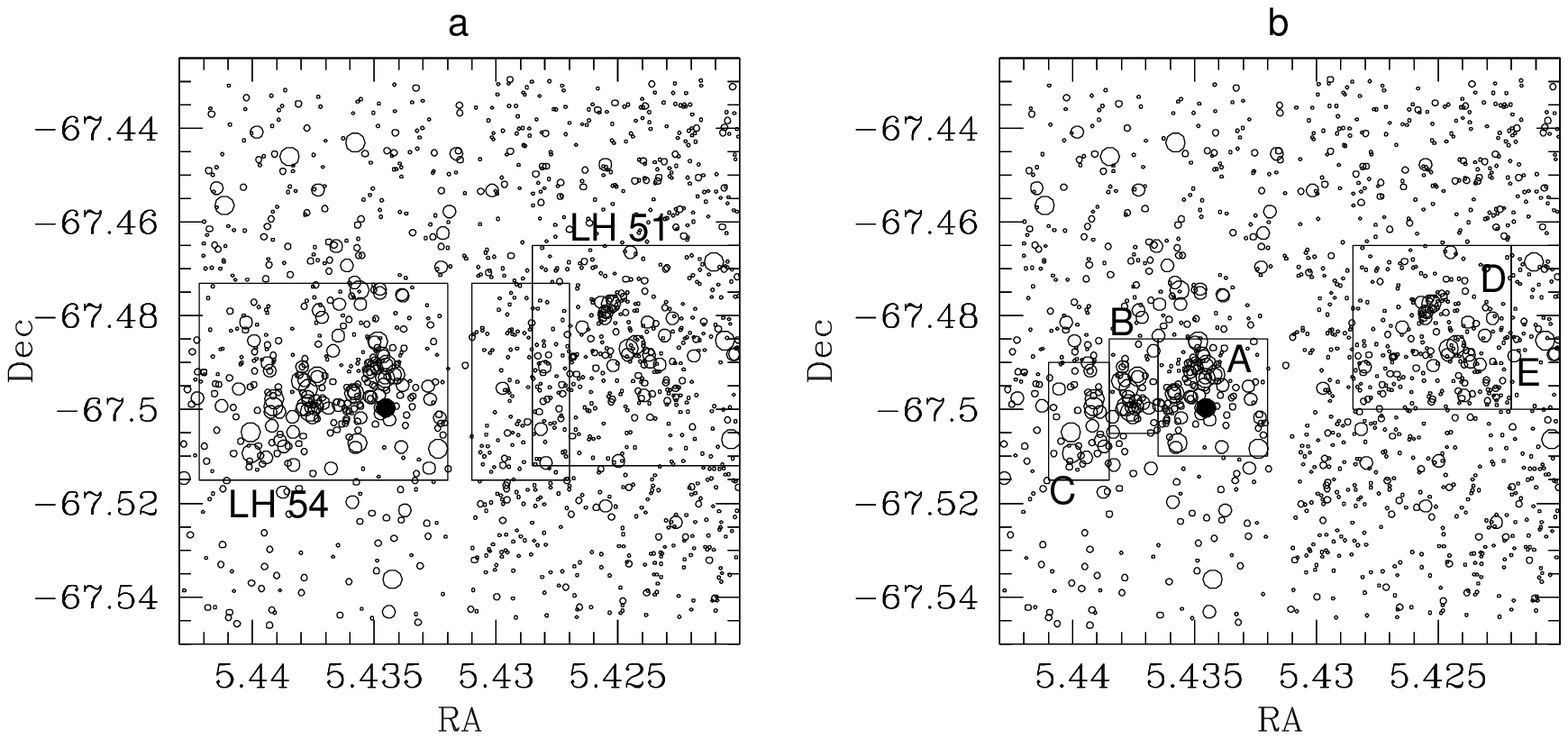}
\end{center}
\caption{Assumed boundaries for ($a$) LH~51 and LH~54; and ($b$)
subregions A -- E.  Point size is assigned
according to $M_{\rm bol}$, with bin separations at $M_{\rm bol}= -7,\
-4,$ and $-1$.  Panel ($a$) also shows the central region studied in
Figure~\ref{hrd.c}$a$.  The WC~5 + O8 Iaf binary star, Sk --67 104, 
is identified by the solid point.
\label{map}}
\end{figure*}

\begin{figure*}
\begin{center}
\epsfbox[20 430 500 600]{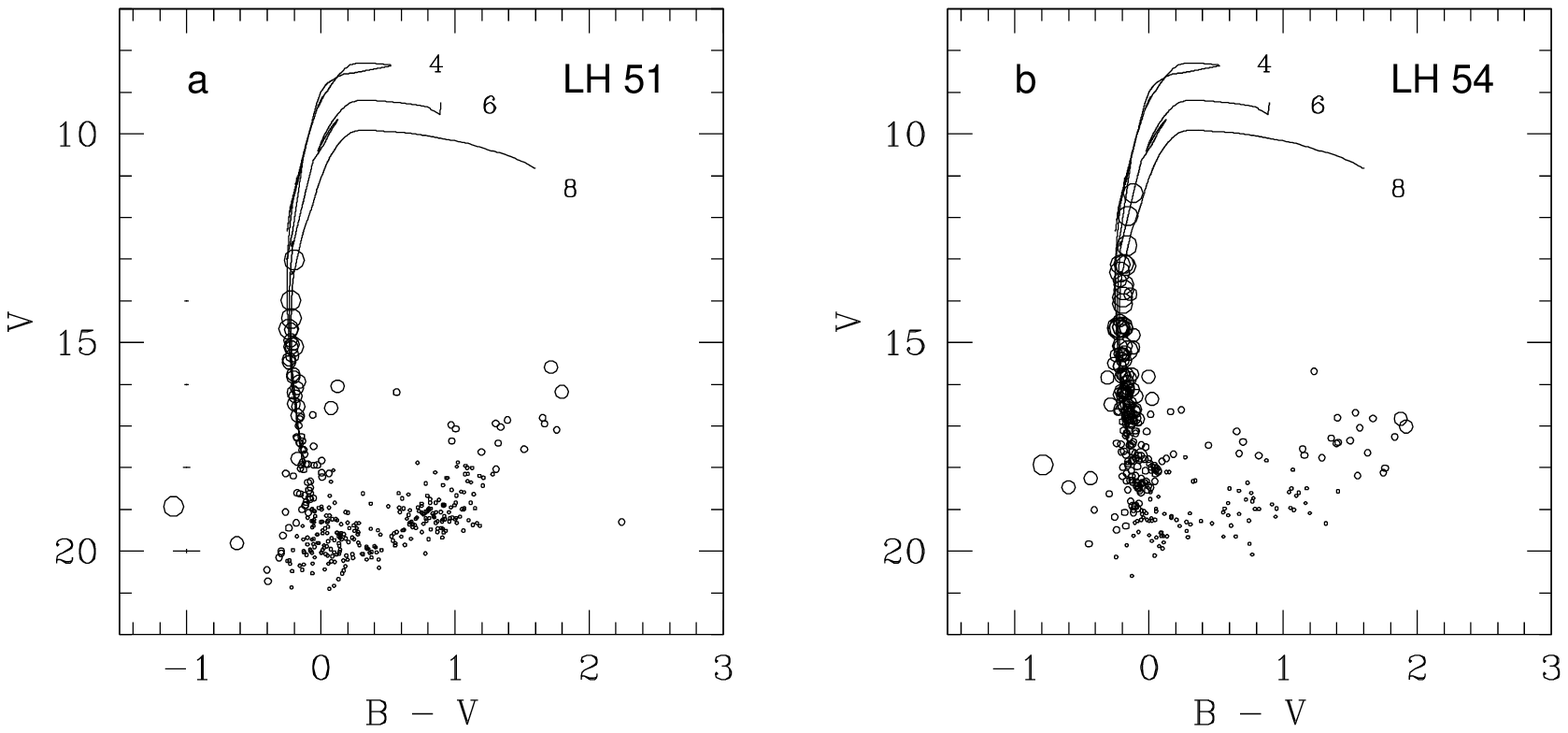}
\end{center}
\caption{CMDs of $V$ vs. $B-V$ for ($a$) LH 51 and ($b$) LH~54, with
isochrones overplotted for 4, 6, and 8 Myr.  Point sizes are assigned
as in Figure~\ref{map}.  Representative photometric error bars are shown
along the left side of Figure~\ref{BV.RL}$a$, at the corresponding $V$ mag.
\label{BV.RL}}
\end{figure*}

In Figure~\ref{BV.RL}$a$ and $b$, we show the color-magnitude diagrams (CMDs)
of $V$ vs. $B-V$ for LH 51 and LH 54, respectively, as delineated by
the boundaries shown in Figure~\ref{map}$a$.  There are four point
sizes, assigned according to the derived bolometric magnitude,
separated by $M_{\rm bol}$ of --7, --4, and --1.  These CMDs show the data
uncorrected for reddening, but with the isochrones adjusted for the
median $E(B-V)=0.08$.  The overplotted isochrones are for 4, 6, and 8
Myr, computed with the code of G. Meynet.  We ran this code with the
stellar models of Schaerer {\etal}(1993), which include convective
overshooting, and adopt a metallicity of $Z=0.008$, appropriate to the
LMC.    

The data are remarkable
in the extremely narrow locus of the association members, suggesting
coeval star formation, low differential reddening, and small
photometric errors.  This of course does not preclude the effect of
systematic uncertainties, which we estimate to be $\lesssim 0.03$ mag
for stars with $V\lesssim 17$,
dominated by transformation errors and aperture corrections (Oey 1996b).
These are included in the error bars shown in Figure~\ref{BV.RL}$a$.
At first glance, the CMDs appear to be well-fitted by
isochrones up to 6 or 8 Myr.  However, it is apparent that only the
very brightest stars are positioned to differentiate between the
isochrones.  
The H-R diagrams we construct below will demonstrate that the
associations are in fact much younger. 

\begin{figure*}
\begin{center}
\epsfbox[30 430 500 620]{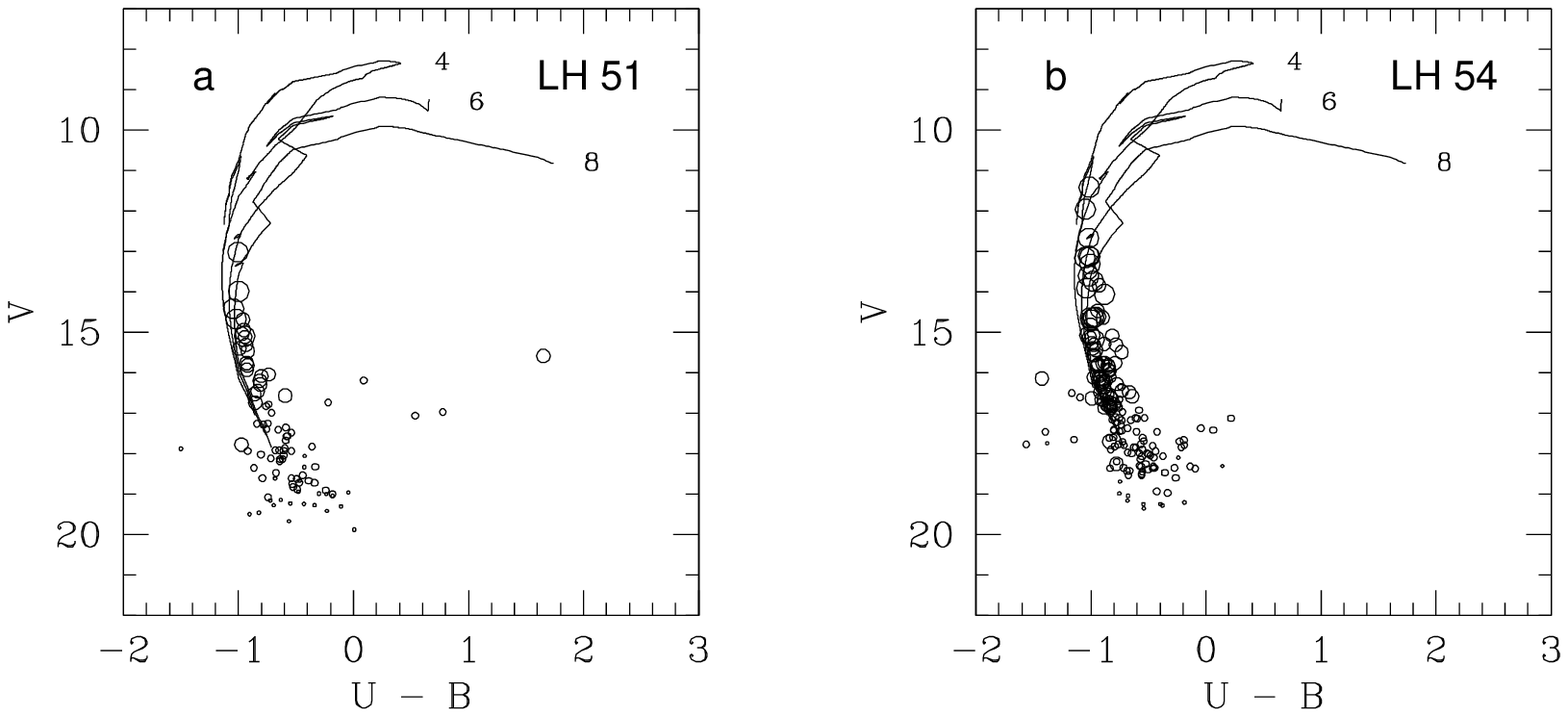}
\end{center}
\caption{CMDs of $V$ vs. $U-B$ for $(a)$ LH~51 and ($b$) LH~54, with
isochrones overplotted.
\label{UB.RL}}
\end{figure*}

It is apparent that the relative unreliability of the $B-V$ CMD stems
from the degeneracy in $B-V$ colors (\eg Massey {\etal}1995).  This
can also be seen from the 
inconsistency in age determinations suggested by the $V$ vs. $U-B$ CMDs
(Figure~\ref{UB.RL}).  Since $U-B$ is more sensitive to $T_{\rm eff}$
for OB stars, the same isochrones diverge more, allowing a more reliable
age determination.  A younger isochrone is implied in
Figure~\ref{UB.RL}, in order to match
the most luminous stars.  However, even in these CMDs one can see that the
hottest stars are also degenerate in their colors.  Furthermore, there
appears to be a problem in the $U-B$ isochrones, which may stem from
the poor constraints on the UV energy distributions of OB stars.  The
isochrones in Figure~\ref{UB.RL} do not match the locus of the stars
in the CMD.  The difference appears to be roughly 0.2 mag in $U-B$,
which is much larger than the typical errors for the stars.  While
transformation errors are largest in $U$, they do not exceed 0.023
mag, the worst uncertainty among the four nights of
observations.  The transformations were computed from observations of
12 to 17 standard stars per night.  The mismatch in observed and
predicted $U-B$ colors has also been reported by other studies
(\eg Will, Bomans, \& Dieball 1997; Hunter {\etal}1995).  Ultimately,
more reliable $U-B$ isochrones will naturally be more useful than
those in $B-V$. 


\begin{figure*}
\begin{center}
\epsfbox[30 430 500 620]{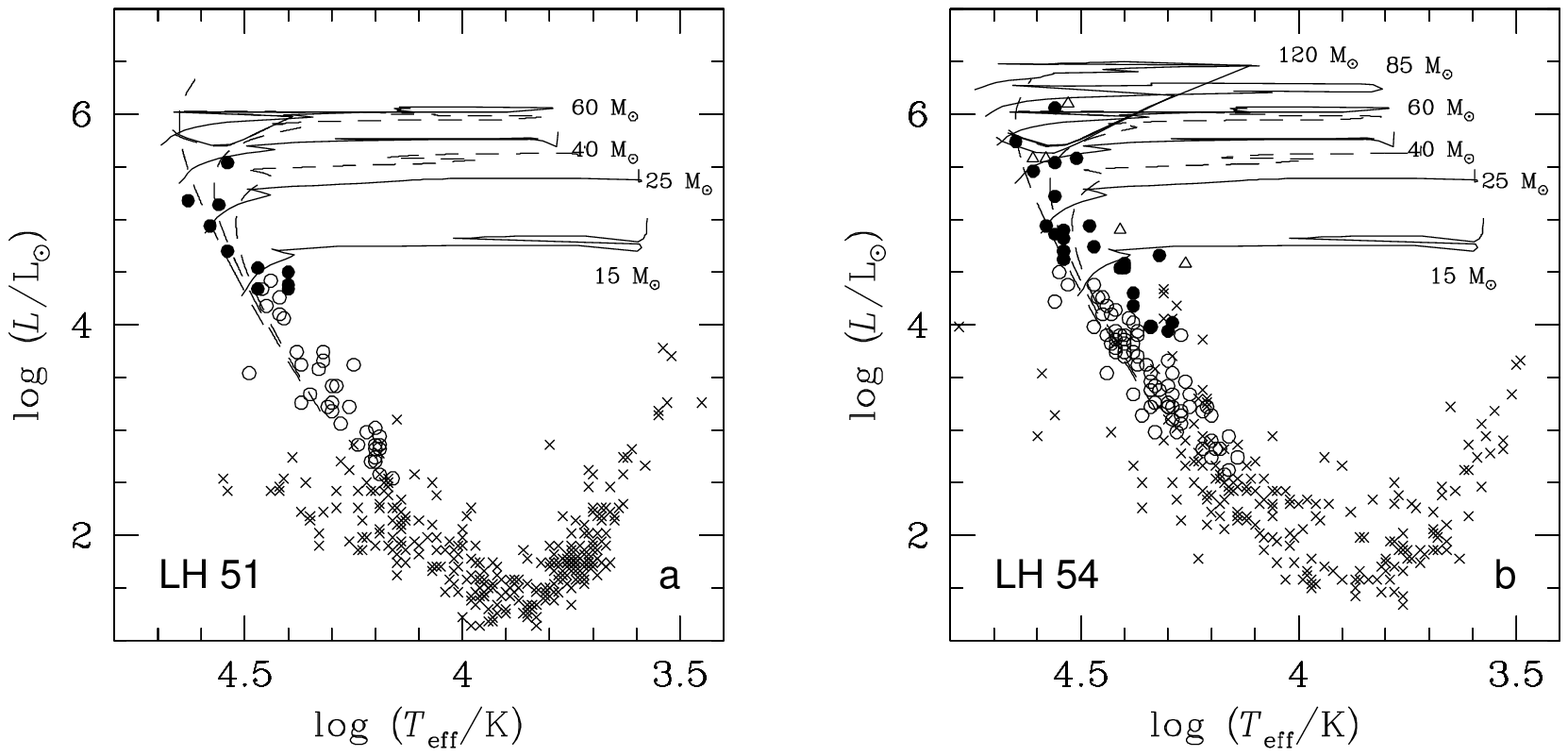}
\end{center}
\caption{H-R diagrams for ($a$) LH~51 and ($b$) LH~54.  Solid points indicate
stars with spectral classifications; open circles show stars placed
using $U-B$ and $B-V$ colors; crosses show those placed from $B-V$
only; and triangles show components of spectroscopic binaries.  The WR
star Br~31 is not shown.  Evolutionary tracks (solid lines) are shown
for stellar models with the masses indicated, and the isochrones
(dashed lines) shown are for 2 (upper), 4 (middle), and 6 Myr (lower).
\label{HRD.RL}}
\end{figure*}

In Figure~\ref{HRD.RL} we show the physical H-R diagram (HRD) for LH 51 and LH
54, with stellar evolutionary tracks overplotted for the indicated
masses (solid lines).  We also show isochrones for 2, 4, and 6 Myr
with the dashed lines.  Stars plotted with solid circles are those
with spectral classifications; open circles show those whose parameters
were derived from both $U-B$ and $B-V$ colors; crosses show those
derived from $B-V$ only; and triangles show components of
spectroscopic binaries.  The Wolf-Rayet (WR) star Br~31 is not included on
the HRDs.  From Figure~\ref{HRD.RL}, the isochrones now show
an age of  $\lesssim 4$ Myr for LH 51, and $\lesssim 3$ Myr for LH 54.
With this age constraint, the presence of the WR star suggests that
LH~54 must be about 3 Myr old (\eg Schaerer {\etal}1993).
There is no evidence of any age spread among or between the
associations, and no evolved stars below the upper-mass turnoff, as is
often seen in other OB associations (Massey {\etal}1995).  The star
formation therefore appears to be coeval to within 2--3 Myr, the
lifetimes of the highest-mass stars.  However, in \S 3.2 below, we
suggest the possibility of an age differential between LH~51 and LH~54
based on evidence from the IMF. 

\begin{figure*}
\begin{center}
\epsfbox[20 180 500 720]{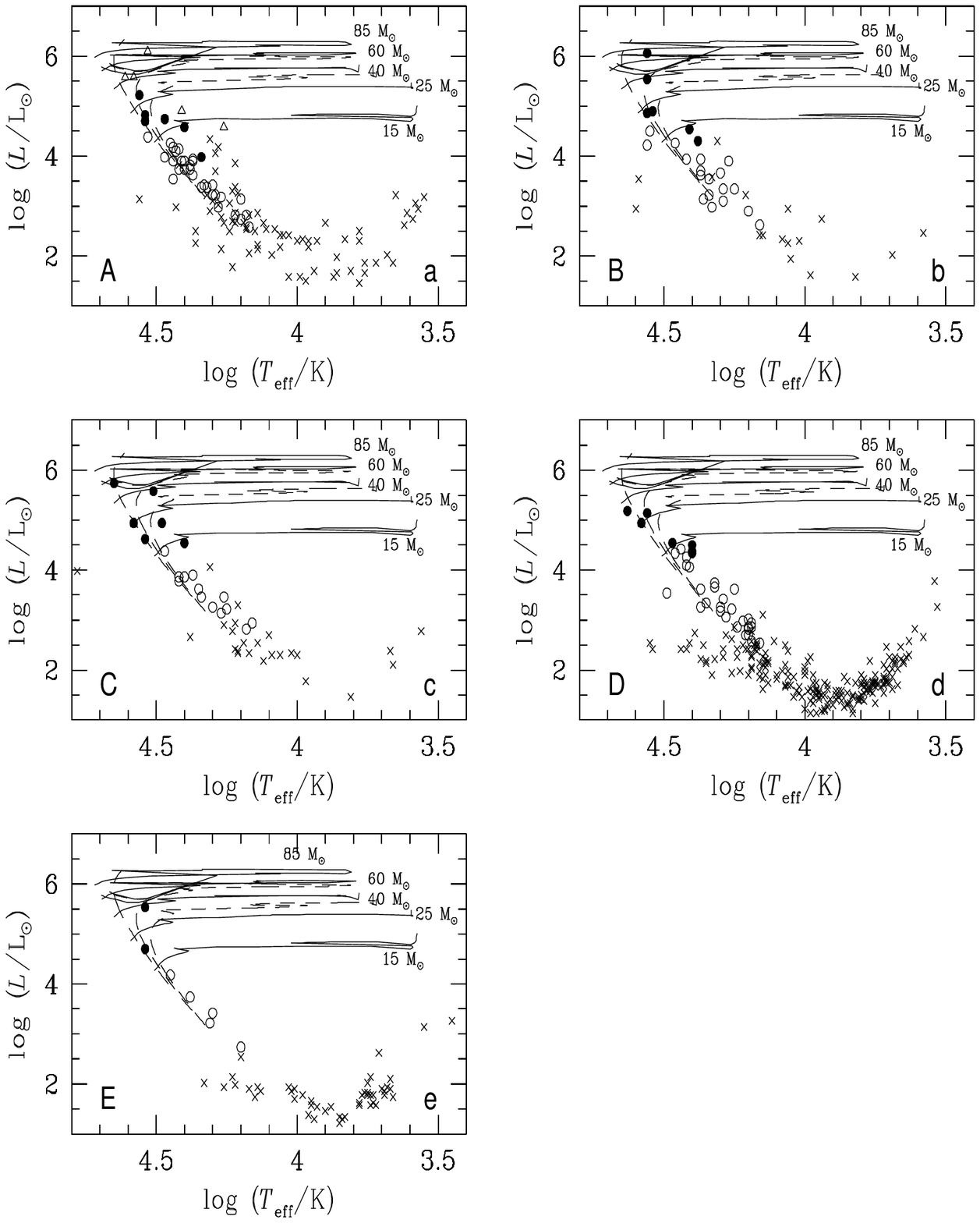}
\end{center}
\caption{H-R diagrams for subregions ($a$) A, ($b$) B, and ($c$) C in
LH~54; and ($d$) D and ($e$) E in LH~51.  Symbols are as in
Figure~\ref{HRD.RL}. 
\label{HRD.sub}}
\end{figure*}

\subsection{Populations in Subregions}

Given this limit for coevality of the star formation, it is therefore
difficult to infer an age gradient in the superbubble.  However,
we can examine different subgroups of stars within the associations,
to search for possible hints revealed by the spatial distribution of
stellar masses.  Figure~\ref{map}$b$ shows the finding chart for
groups A, B, and C in LH 54; and groups D and E in LH 51.  The HRDs
for these regions are shown in Figure~\ref{HRD.sub}.  The
subgroups in LH 54 all show high-mass stars around $60 \msol$ or
above, and regions D and E in LH 51 both show stars near $40 \msol$.
Thus the spatial distribution of masses appears to be highly uniform.
The age differentials again appear to be no greater than $\sim2$--3
Myr, and could well be less.  These OB associations therefore do not
internally exhibit propagating star 
formation on any greater timescales.  This is in marked
contrast to star forming complexes like DEM 34, which, like DEM 192,
has an associated superbubble, but shows clear age
differences among different subgroups (Walborn \& Parker 1992).

If the stars were truly coeval, then what would be the origin of the
superbubble? 
It is peculiar that the stars are concentrated toward the edges of the
shell, rather than within the center.  If LH 51 and LH 54 were born
before the superbubble, then presumably we would expect two shells
centered on the associations.  This is clearly not the case.  We
suspect that pre-existing geometry of the gas must play a role in the
position of the shell, but it is difficult to believe that two
separate gas clouds have conspired to create an almost perfectly
spherical shell, with fairly well-behaved expansion kinematics
(Lasker 1980; Meaburn \& Terrett 1980).

One clue may lie in the fact that the western edge of LH 54 is located
closer toward the center of the superbubble, where the gas appears
relatively cleared out (Figure~1).  It may
therefore be that the stars on the west side of LH 54 were surrounded
by less gas from the natal cloud, therefore initiating the
superbubble's formation around this region.  The position of the stars
in LH 51 suggests that they have not contributed substantially to the
formation of the superbubble.  This may be because they
younger, within the age constraints found above.  We note that 
LH~51 contains no evolved stars whatsoever, whereas LH~54 contains the
WR star Br~31.  LH 51 also contains fewer
stars, and somewhat lower-mass stars, as is expected from the
stochastic nature of the highest-mass star formation in sparser clusters.
This association therefore has less stellar wind power to disperse its
surrounding gas, although we note that the reddening values for LH~51 actually
appear to be slightly lower than those for LH~54.

\begin{figure*}
\begin{center}
\epsfbox[30 430 500 620]{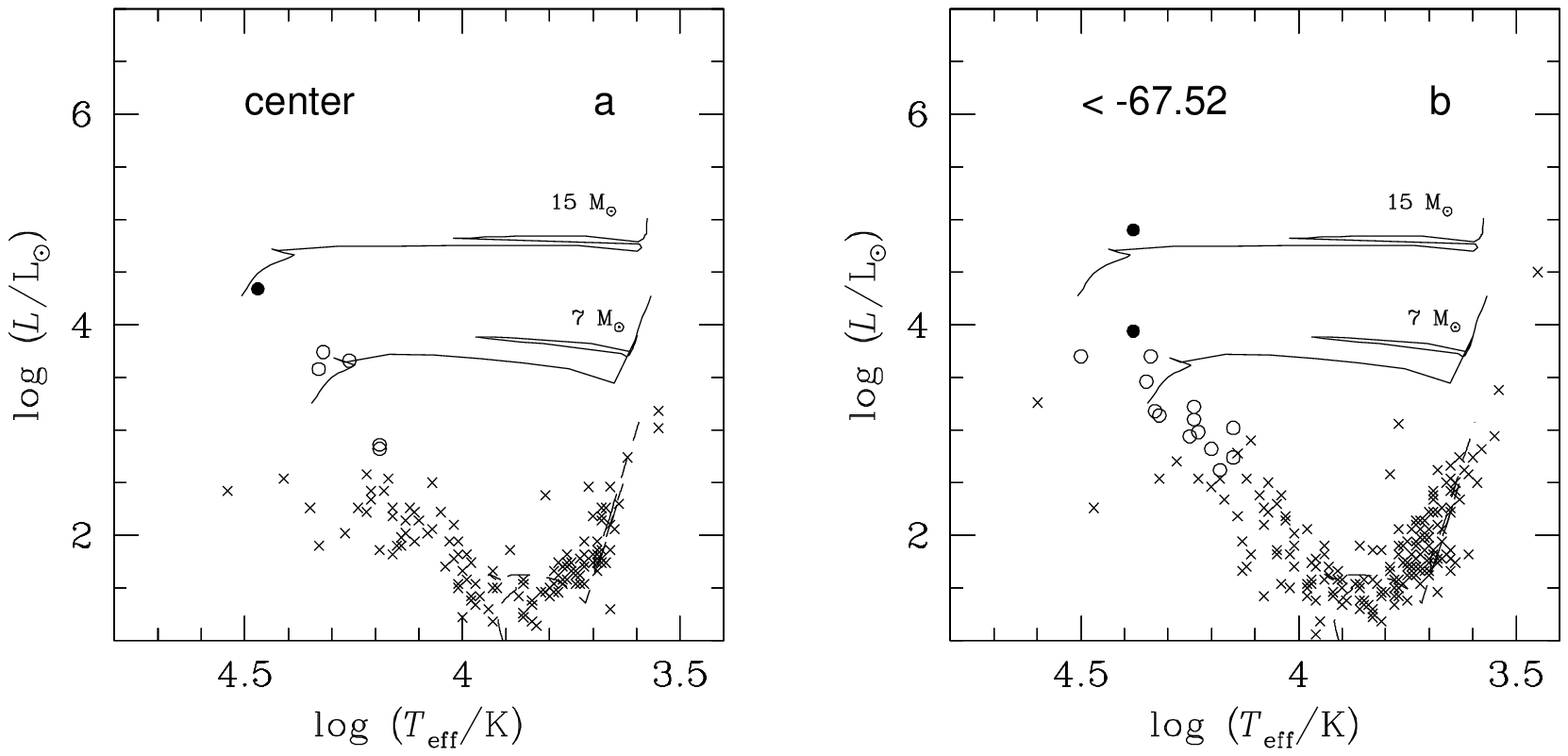}
\end{center}
\caption{H-R diagrams for ($a$) the central region defined in
Figure~\ref{map}$a$, and ($b$) the region south of $-67.52^\circ$.
Evolutionary tracks for stars of 7 and 15 $\msol$ are shown as
indicated (solid lines), and the dashed line shows the isochrone for 1
Gyr.  Symbols are as in Figure~\ref{HRD.RL}.
\label{hrd.c}}
\end{figure*}

Could it be that a previous stellar population, more closely centered
within the superbubble, is the origin of the shell?  In
Figure~\ref{hrd.c}$a$, we show the HRD for stars in the central region
between LH~51 and LH~54, delineated in Figure~\ref{map}$a$.
Figure~\ref{hrd.c}$b$ shows the HRD for all 
stars south of $-67.52^\circ$, which, lying on the outskirts of the
region, should resemble the background population.  We do caution that
most of this region still appears to be within the confines of the
superbubble in projection, but given the fairly concentrated
distribution of stars in the OB associations, this area is likely to yield
characteristics of the background.  There appears to be no significant
difference in the HRDs in Figure~\ref{hrd.c}.  The highest-mass stars in 
each population ($7 - 15\ \msol$) could be members of the OB
associations, so the age determinations are not well-constrained.
However, both the central region and outskirts suggest ages of 
several$\times 10^7$ yr, with an older, underlying background of red giants
with an age of order $\sim 1$ Gyr.  This is
consistent with the burst in field star formation beginning
around 2 Gyr ago (\eg Vallenari {\etal}1996).  Given that the younger
populations in Figure~\ref{hrd.c} appear to be similar both in the center
and outskirts of the shell, we infer a continuous background with this
age of several$\times 10^7$ yr, which is therefore unlikely to have
spawned the superbubble.

\subsection{The Initial Mass Function}

We compute the IMF for the OB associations following Massey
{\etal}(1989), where $\xi(\log m)$ is defined as the number of stars
per logarithmic mass interval per kpc$^{2}$.  Assuming coeval star
formation as suggested above, the present-day mass function
reflects the IMF.  The evolutionary tracks
for the different masses are used to define the mass bins for the
stellar census (Figure~\ref{HRD.RL}), where the tracks define the bin
limits.  We then correct for the bin size and spatial area for each
region defined in Figure~\ref{map} to obtain $\xi(\log m)$, for all
stars with masses $> 7\ \msol$.  We computed the slope $\Gamma$ with
a least-square polynomial fit, weighted by root-$N$ in each bin.

\begin{figure*}
\begin{center}
\epsfbox[60 540 500 630]{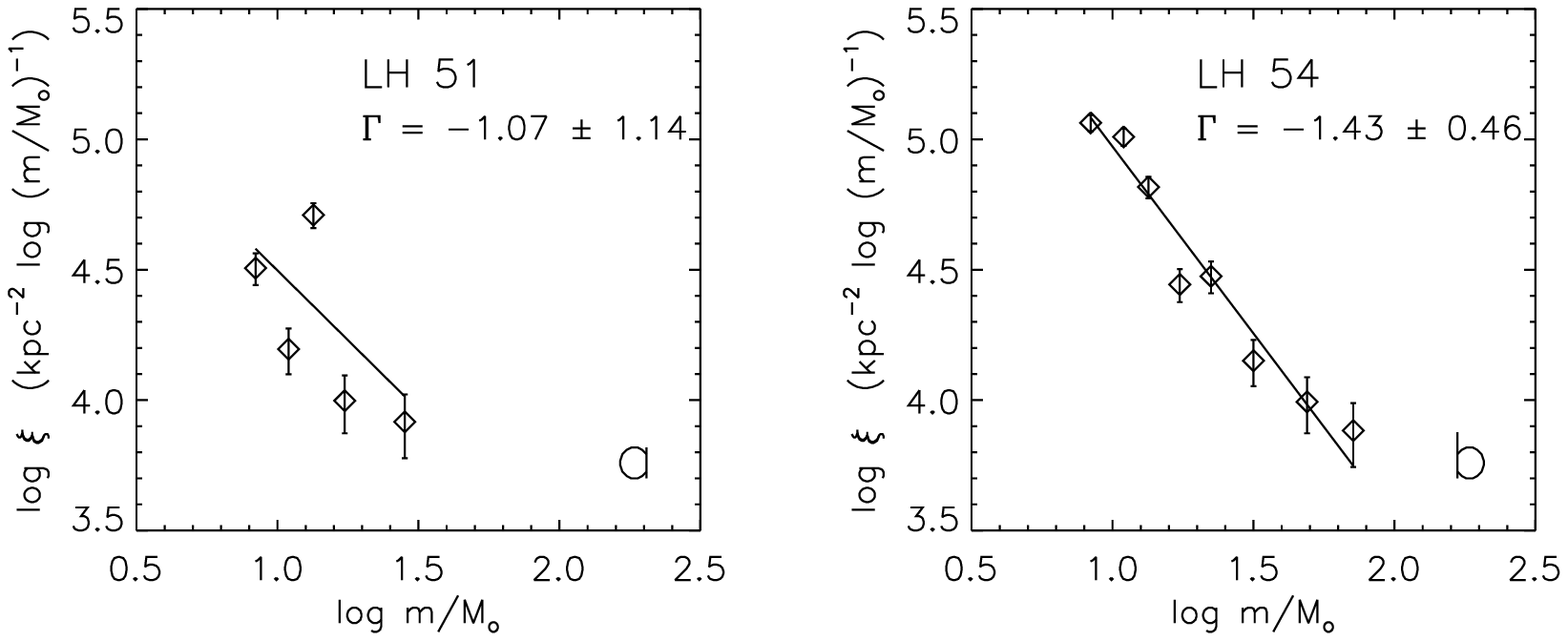}
\end{center}
\caption{The IMF for ($a$) LH~51, and ($b$) LH~54.  For reference, the
Salpeter slope is $\Gamma= -1.35$.
\label{imf}}
\end{figure*}

In Figure~\ref{imf} we show the IMF for LH~51 and LH 54.  LH~51 shows
an excessively large number of stars between 12 and 15 $\msol$, but
the IMF for both associations agree, within the errors, with the
Salpeter (1955) value:  the slope $\Gamma = -1.07\pm 1.14$ and
$-1.43\pm 0.46$ for LH~51 and LH~54, respectively.  For comparison,
the Salpeter value is $\Gamma = -1.35$ in the $\log-\log$ space.

We also briefly examined subregions A -- E to check for variations in
the slope of the IMF.  For regions A, B, C, and D, we found $\Gamma =
-1.53=\pm 0.69,\ -1.33 \pm 1.05,\ -1.32\pm 1.04,$ and $-1.19\pm 1.23$,
respectively.  In region E, no meaningful fit can be obtained since
there are only 4 stars with masses $> 7\ \msol$.  It is again apparent
that there is no noticeable variation among the subregions.  The small
numbers of stars in each subregion cause the uncertainties in the
slope fit to be large, but it is apparent from the values of $\Gamma$
that the IMF remains broadly consistent with the Salpeter value on smaller
spatial scales within the associations. 

In LH~51, the anomalously large number of stars in the 12--15$\msol$ bin is
intriguing:  could this represent the overlap between the underlying
field star mass function and the lower-mass limit of the extremely
young OB association?  The isochrones for 1 and 2 Myr have lower-mass
limits of 12 and 20$\msol$, respectively.  The field star mass
function appears to have a present-day upper-mass limit in the range
7--15$\msol$, as inferred from the outlying regions examined above
(Figure~\ref{hrd.c}).  These ages and 
masses are therefore consistent with the interpretation of an overlap
between the two mass functions. Since a similar excess feature in the
mass function is not observed in LH~54, this supports the older age of
3 Myr, inferred above for this association.  For reference, a 3 Myr
isochrone has a lower-mass limit of 9 $\msol$.  We note that field
star contamination may therefore be present in the IMF computation for
LH~54 as well, resulting in slight overpopulation in the 1--2 lowest bins
shown in Figure~\ref{imf}$a$.  With this hint of a younger age for LH~51,
it thus might be possible that the superbubble
was created by LH~54, which may then have triggered the creation of LH~51.

\section{Superbubble Modeling}

The evolution of the superbubble itself can shed light on its
relationship to the OB associations.  To study the dynamics of this
shell, we use the same code described by Oey \& Massey (1995) and
Oey (1996a).  This code numerically integrates the equations of motion
for an adiabatic bubble with a thin shell (\eg Weaver {\etal}1977; Ostriker
\& McKee 1988), allowing variable input wind power.  We have seen in
the previous section that the stars in LH 51 are lower mass and at
shell's western edge.  These factors indicate that LH 51 is
unimportant as a contributor to the growth of the superbubble.  In
fact, the most 
intrinsically luminous star in LH 51 is L51N-1, which appears to be
located on the external side of the shell (Oey 1996b).  We therefore
conclude that LH 54 is primarily responsible for the superbubble.

As shown by Oey \& Massey (1995) and Oey (1996a), it is the very most
massive stars that dominate shell growth.  We will consider the
observed stars in LH 54 with masses $\geq 40 \msol$.  The masses are
assigned as the nearest evolutionary track in the HRD.  As in our
previous work, we estimate the mechanical wind power $L_w$ based on the stellar
evolution models of Schaerer (1993) for LMC metallicity.  These assume
the stellar mass-loss rates ($\Mdot$) parameterized by de~Jager,
Niewenhuijzen, \& van der Hucht (1988; hereafter JNH88), scaled for
LMC metallicity.  However, here, we reexamine the formulation for
computing the expected power $L_w$ for the H-burning main sequence.  

\subsection{What is $\Mdot(t)$?}

\begin{figure*}
\begin{center}
\epsfbox[-50 540 500 700]{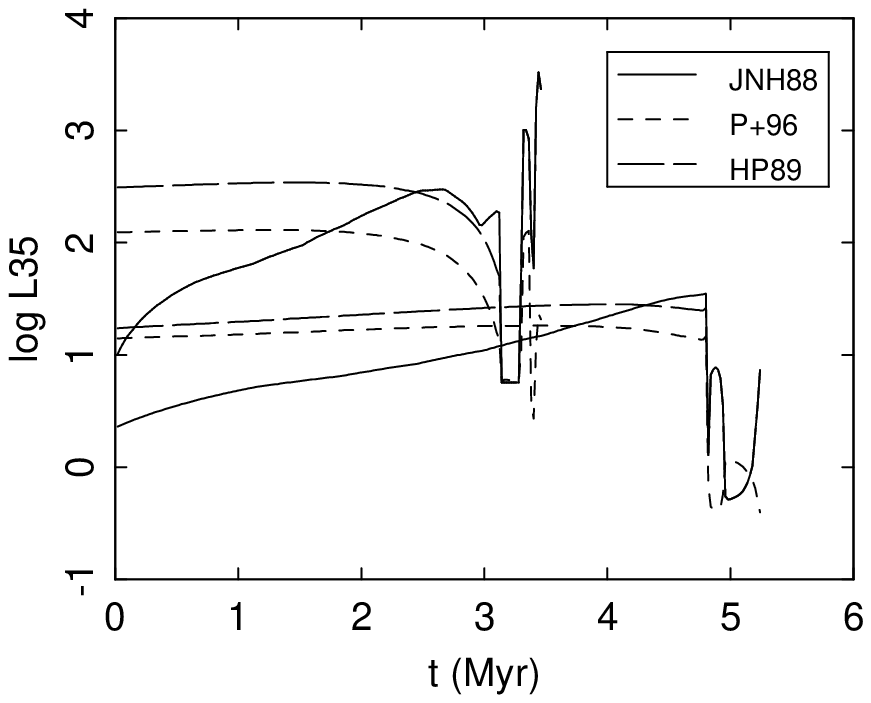}
\end{center}
\caption{Comparison of $L_w(t)$ in units of $\log L_w/(10^{35}\
\ergs)$, estimated from the $\Mdot$ parameterizations of JNH88 (solid
line), P+96 (long-dashed line), and HP89 (short-dashed line).  The
upper set of models is for an 85 $\msol$ star, and the lower set is
for a 40 $\msol$ star.  Beware that $L_w(t)$ for P+96 and HP89 are not
self-consistent with the stellar evolution timescales (see text).
\label{Lcompare}}
\end{figure*}

In our earlier work, we estimate the wind power by,
\begin{equation}\label{Lw}
L_w = \frac{1}{2}\Mdot v_\infty^2 \quad ,
\end{equation}
where $\Mdot$ is from JNH88, as used to compute the
stellar models.  We obtain the stellar wind terminal velocity
($v_\infty$) from the escape velocity, as given by the empirical
parameterization of Howarth \& Prinja (1989, equations~10--11,
hereafter HP89).  The $L_w(t)$
resulting from this formulation is shown in Figure~\ref{Lcompare} for
85 and 40 $\msol$ models (solid line).  We now consider 
the wind momentum -- stellar luminosity relation determined by Puls
{\etal}(1996, hereafter P+96), using the empirical calibration,  
\begin{equation}\label{puls}
\log(\Mdot v_\infty) = 1.49 \log\biggl(\frac{L_*}{\Lsol}\biggr) - 
	\log\biggl(\frac{R_*}{\Rsol}\biggr)^{1/2} + 20.26 \quad ,
\end{equation}
derived for luminosity class II - V (J. Puls, private communication).
$L_*$ and $R_*$ denote the stellar luminosity and radius, respectively.
We then estimate $L_w(t)$ from equation~\ref{Lw}, using $v_\infty$
obtained as before.  The resulting wind power is shown by the
long-dashed line in Figure~\ref{Lcompare}.  Finally, we also compute
$L_w(t)$ using the empirical parameterization for $\Mdot$ of HP89 (their
equation~18).  This result is shown by the short-dashed
line in Figure~\ref{Lcompare}.

Since $v_\infty$ is computed the same way in all three recipes for
$L_w(t)$, the comparison in Figure~\ref{Lcompare} essentially reflects
the different behavior in $\Mdot(t)$.
It is apparent that the $\Mdot$ relations of JNH88
show a different evolution than those of P+96 and HP89.  $L_w(t)$ for
P+96 and HP89 are similar in form to each other, showing fairly constant wind
power over the H-burning period.  In contrast, $L_w(t)$ for JNH88
increases by an order of magnitude or more over the same period,
with initially much lower $L_w$ than those for P+96 and HP89.
Castor (1993) points out that the discrepancy
of the JNH88 parameterization is caused by fitting the relations 
without the benefit of recent understanding of massive star evolution.
Hence, it is desireable to use the more updated relations of P+96 or HP89.

However, we caution that the stellar models assume the $\Mdot$
relations of JNH88, and therefore the $L_w(t)$ plotted for P+96 and HP89
are not self-consistent with the modeled stellar evolution.
We may expect that the evolution would progress much more rapidly if
the higher mass-loss rates of P+96 or HP89 are incorporated in the 
models.  The net result is that the H-burning phase might be
shortened, as would our estimated $L_w(t)$ in Figure~\ref{Lcompare}.
It would be useful to constrain stellar models with independent empirical
tests for the lifetimes of the most massive stars.  

\subsection{Models for DEM 192}

Given the caveats for the estimates of $L_w(t)$,
we now proceed to model the superbubble DEM 192.  We estimate the rms
electron density $n_e$, and thereby the H density, of the shell from the
\Ha\ emission measure, as described by Oey (1996a).
Consideration of both center and edge lines of sight yield $n_e\sim 8\
\rm cm^{-3}$.  For an originally uniform and homogenenous ambient
density, the swept-up material in the shell represents an 
initial ambient density of $n \sim 2\ \rm cm^{-3}$.  Studies of the
shell kinematics by Lasker (1980) and Meaburn \& Terrett
(1980) indicate an expansion velocity $v_s \sim 35 \kms$.

In LH 54, there are two 85 $\msol$ stars, one 60 $\msol$ star, 
and five 40 $\msol$ stars (Figure~\ref{HRD.RL}).  In addition, there
is also the WC~5 star, Br~31.  If we assume coeval star formation,
stellar evolution predicts that that this star evolved from a
progenitor more massive than any of its currently H-burning siblings.
In addition, DEM~192 also shows enhanced X-ray emission
(Chu \& Mac Low 1990; Wang \& Helfand 1991), which is likely to be
the signature of a recent supernova remnant (SNR) impact.  The large shell 
velocity also suggests that it belongs to the category of objects
whose shells have been accelerated, most likely by SNR activity (Oey
1996a; see below).  Indeed, the IMF for LH~54 derived in \S
3.2 predicts that one additional star with mass in the range 85--120
$\msol$ should be expected, which would correspond to a supernova (SN)
progenitor.  We assign a mass of 120 $\msol$ for the SN progenitor, and 85
$\msol$ to the progenitor of Br~31.  This yields a model stellar
population of one 120, three 85, one 60, and five 40 $\msol$ stars.

\begin{figure*}
\epsfbox[30 460 500 660]{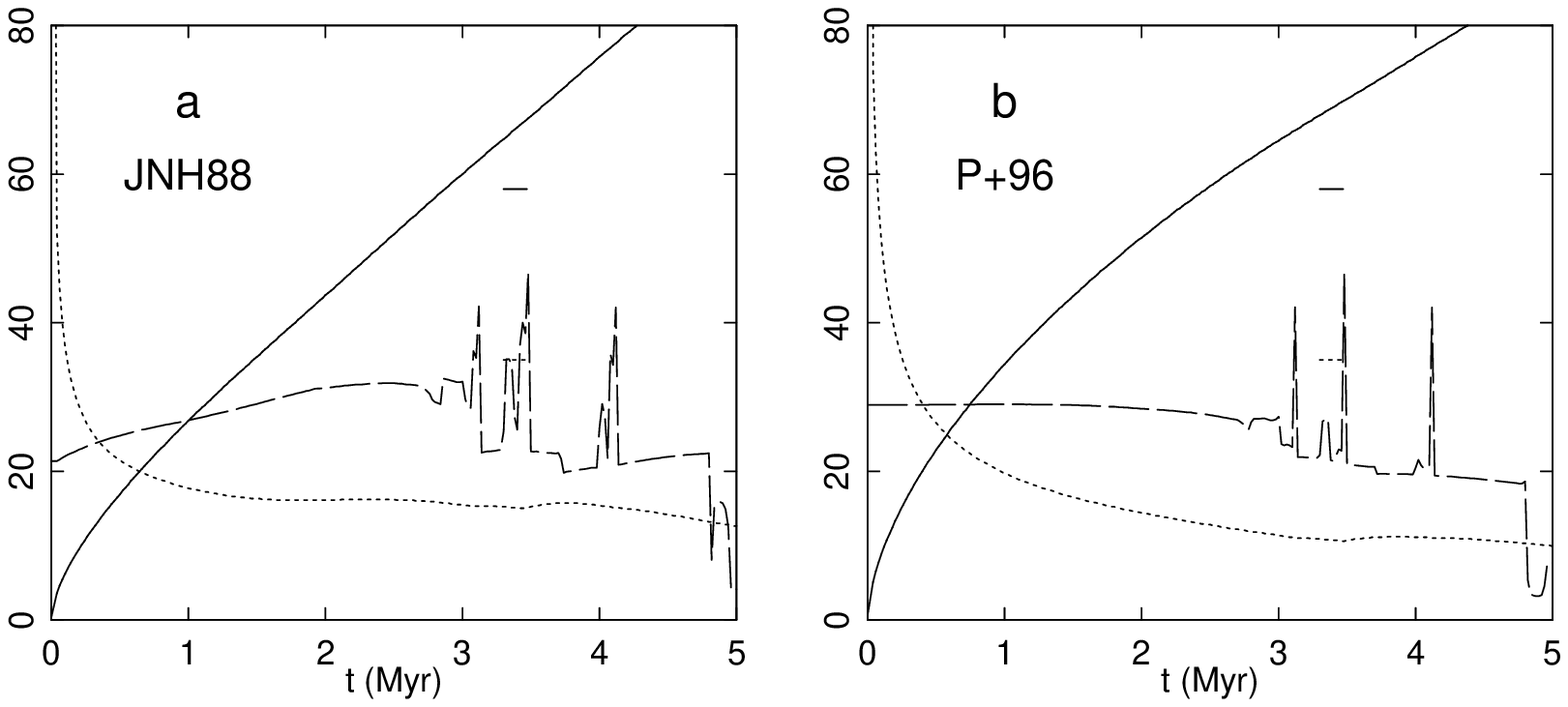}
\caption{Evolution of shell parameters for the coeval creation of
LH~54:  one 120, three 85, one 60, and five 40 $\msol$ stars.  The
evolution of the shell radius $R_s$/pc (solid line), expansion velocity
$v_s/\kms$ (dotted line), and $10\log(L_w/10^{35}\ \ergs)$
(dashed line) are shown, for $\Mdot$ given by JNH88 ($a$) and P+96
($b$).  Note that $R_s$ and $v_s$ are shown for $0.1\times L_w/n$,
although $L_w$ shown is not adjusted.  The short, horizontal solid and
dotted lines indicate the observed shell radius and expansion velocity
for DEM~192, over the predicted age range suggested by the presence of
the WR star.
\label{mod.3}}
\end{figure*}

Figure~\ref{mod.3} shows the models for the superbubble evolution
resulting from the combined mechanical power of these stars in LH 54 
with masses $\geq 40 \msol$, assuming coeval star formation.
The input wind power is shown by the
dashed line, in units of $10 \log (L_w/10^{35} \ergs)$, as a
function of time $t$ in Myr.  Figure~\ref{mod.3}$a$ shows the model
for $L_w(t)$ estimated as in our previous work, based on $\Mdot$ of JNH88;
and Figure~\ref{mod.3}$b$ shows the model for
$L_w(t)$ estimated from the momentum-luminosity relation of the Munich
group (equation~\ref{puls}).  We refer the reader to Oey \& Massey
(1995) for details concerning evolution beyond the H-burning phase; as
can be seen in that work, and Oey (1996a), these stages are
dynamically unimportant for 
young superbubbles $\lesssim 4$ Myr old.  The brief WR and SN
stages can be seen in the form of $\log L_w(t)$, however, where the WR phase
corresponds to the double-peaked structure preceding the final SN spike.

The solid line in Figure~\ref{mod.3} shows the growth of the
superbubble radius ($R_s$) in pc, and the dotted line shows the
evolution of its expansion velocity ($v_s$) in $\kms$.  The observed
superbubble radius of 58 pc and observed expansion velocity of $35
\kms$ are shown by the horizontal solid and dotted lines,
respectively.  We use the extent of the
horizontal lines representing the observed parameters, to indicate the
``observation window,'' which is the range in age that is consistent
with the observed stellar population.  For the model to agree with
observation therefore requires that the solid lines and short-dashed
lines intersect at the same age $t$.  For DEM~192, the presence of the
WR star severely constrains the age of the shell for coeval star
formation, since this stellar phase is extremely short-lived.  
Figure~\ref{mod.3} therefore shows a short observation window during
the WR phase of Br~31.

The study of superbubble dynamics by Oey (1996a) implied a growth-rate
discrepancy, also suggested by previous authors (\eg Garc\'\i a-Segura
\& Mac Low 1995; Drissen {\etal}1995), of up to a factor of 10
overestimate in the implied $L_w/n$ for the standard evolution.  For our
model of DEM 192, we therefore take $0.1\times L_w/n$ of the observed stellar
and ambient parameters.  Figure~\ref{mod.3}$a$ shows that this produces
reasonable agreement in the predicted and observed shell radius $R_s$,
considering the uncertainties in $L_w$ and structure of the ambient
medium (see \S 4.3), as well as the strong age constraint for the
shell.  Similar agreement, with less stringent age constraints, was
found for virtually all of the seven objects studied by Oey (1996a)
and Oey \& Massey (1995).  We therefore find that the shell kinematics
are fully consistent with an origin from the stellar winds of LH~54,
which also supports the age estimate of $\sim 3$ Myr for the system.
This again refutes the existence of an earlier stellar population for
creating DEM~192.  

Can the more realistic representation of $\Mdot$ given by P+96 and
HP89, discussed above, resolve the growth-rate discrepancy for the
superbubbles? 
The more constant evolution for $L_w(t)$ in Figure~\ref{mod.3}$b$
produces a shell that initially grows faster.  However, if the
H-burning phase is significantly shortened, this model is likely to
yield $R_s$ that is the same size, or even smaller than, the shell
size during the equivalent WR stage in Figure~\ref{mod.3}$a$.  It is
thus conceivable 
that a more realistic representation of $\Mdot$ in the stellar models
might help reconcile the growth-rate discrepancy for the superbubbles.
However it appears unlikely that the problem would be resolved altogether.

Oey (1996a) identified two categories of objects:
``low-velocity'' superbubbles, whose kinematics are consistent with
the standard, adiabatic shell growth, aside from the growth-rate
discrepancy; and ``high-velocity'' objects, whose observed expansion velocities
were too high relative to their shell radii to be consistent with the
standard evolution.  These objects also showed enhanced X-ray
emission, hence the anomalous dynamics were suggested to result from
SNR impacts on the shell walls.  Given that the size of DEM 192 is
similar to the 
objects examined in our earlier work, and that the observed expansion
velocity is $\gtrsim 25\ \kms$, we would suspect that this system 
falls in the ``high-velocity'' category.  Indeed, Figure~\ref{mod.3}
shows that, while it is possible to match the observed shell radius
close to the observation window, the modeled $v_s$
is too low, and cannot be matched simultaneously.  As mentioned
earlier, DEM 192 also shows excess X-ray emission, in keeping with
with the other superbubbles in this category.  Our assumption of a SN
progenitor in the parent stellar population therefore appears to be
justified. 

\subsection{Shell acceleration}

As described by Oey (1996a), the elevated $v_s$ for the
high-velocity superbubbles is energetically consistent with 
acceleration by SNR impacts.  While the excess X-ray
emission seen in all these objects also suggests SN activity, here we
also explore an alternative explanation:  a sudden drop in ambient
density.  Oey \& Massey (1995) show the effect of an exponential radial
decrease in ambient density from the center.  Although such a model can
accelerate the shell velocities to the observed values, it cannot
simultaneously match the observed shell radius, since $R_s$ grows far
too rapidly.  However, it is possible that a sudden, discontinuous drop in
ambient density could create a ``blowout'' situation that 
better matches the observations.  This situation may apply to
many objects, as the shells break out of parent molecular clouds.

\begin{figure*}
\begin{center}
\epsfbox[-50 540 500 700]{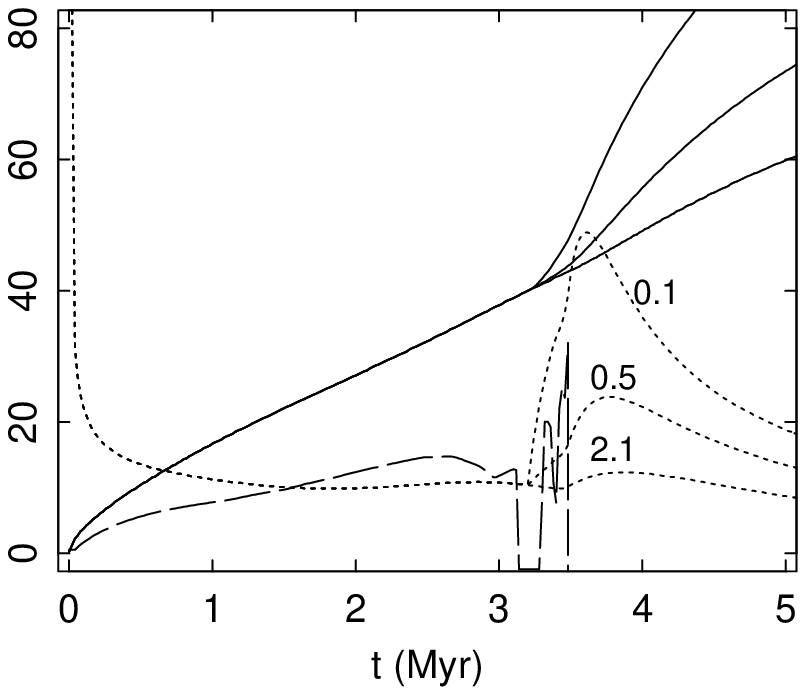}
\end{center}
\caption{Bubble model for an 85 $\msol$ star expanding into an
initially higher density medium, then dropping to a lower-density
environment.  The initial ambient density $n=2.1$ cm$^{-3}$, and models for
a density drop at $R_s=40$ pc to $n=0.5$ and 0.1 are shown, 
along with the model for a continuous $n$.  Line types and units are as in
Figure~\ref{mod.3}. 
\label{density}}
\end{figure*}

Figure~\ref{density} shows the bubble model for an 85
$\msol$ star expanding into an initially uniform density.  To avoid
confusion, we 
assume here that the growth-rate discrepancy is caused entirely by an
overestimate in $L_w$, rather than the combination $L_w/n$.  $L_w$ in
Figure~\ref{density} is therefore reduced by a factor of 0.1.
This implies that the ambient density for DEM 192 is 2.1 cm$^{-3}$,
which we take as the initial, uniform environment for the growing shell.
We assume that at a radius of 40 pc, the density instantaneously drops
to a lower value.  Figure~\ref{density} shows the second zone at
$n=0.5$ and 0.1 cm$^{-3}$, as well as continuing at the original value.

It is apparent that a sudden, discontinuous drop in ambient density 
quickly accelerates the shell expansion velocity before the shell
radius has time to respond.  This results in much lower ratios of
$R_s/v_s$, and is capable of reproducing values observed in the
high-velocity category of superbubbles.  For example, a drop to values
in the range $0.1n$ to $0.3n$ can easily reproduce the observed
parameters for DEM 192.  A spherically uniform drop in density
modeled here could correspond to spherical symmetry found for a star
in an isolated cloud.  However, a more realistic and commonplace
situation is unlikely to present perfectly spherical symmetry, in
which case the blowout acceleration will be localized and even more
pronounced.  Our models therefore represent an upper limit in the density
differential required to match the observations.
While the observed X-ray enhancements do suggest SNR
acceleration for DEM 192, it is apparent that a
low-density blowout can beautifully mimic the shell kinematics.  Such
a situation is furthermore likely to apply to other superbubbles.

\section{Conclusion}

The positions of the OB associations, LH 51 and LH 54, close to the edges
of the superbubble DEM~192, suggest that their star formation was
triggered by the expansion of the shell.  However, the results of our
investigation do not entirely support this scenario.  
There is no evidence of an earlier stellar population to initiate
the shell growth.  Instead, the stellar populations and superbubble
modeling suggest that the stellar winds of the most massive stars
in LH 54 are responsible for creating the superbubble.  The modeled
kinematics of 
DEM~192 are fully consistent with those of the ``high-velocity'' category of
superbubbles examined by Oey (1996a), that show evidence of shell
acceleration by SNR impacts.  The inferred SN progenitor
whose wind was probably significant in creating the shell,
may have been located closer to the center than the remainder of
LH~54, which could explain the puzzling location of the shell center
with respect to the OB associations.  The WC~5 + O8~Iaf binary
Sk~--67~104 (Br~31; L54SA-1) is also located closer 
to the center than most of the other cluster members
(Figure~\ref{map}), lending plausibility to this scenario.  It is also
likely that the initial distribution of the ambient gas played an
important role in the location of the superbubble with respect to the
stars.  We find
that LH~51, on the other hand, is too poor to contribute significantly
to the growth of DEM~192.

OB membership within each association appears to be 
coeval at least to within the H-burning lifetime
of the most massive stars, $\lesssim 2-3$ Myr.  There are no 
evolved stars below the high-mass turnoff, as is often seen in other
OB associations of the Magellanic Clouds (Massey {\etal}1995).  
However, for LH~51, an excess number of stars in the range
12--15$\msol$ might represent the overlap between the
lower-mass limit of the OB association and the upper-mass limit of the
field star mass function.  This therefore may be evidence that LH~51
has an age of 1--2 Myr, and could indeed have been triggered by the
action of the superbubble, whose age must match that of LH~54, roughly
3 Myr old.  

Investigation of five spatial subgroups shows that
the stellar masses appear to be uniformly distributed within each
association, and are again consistent
with coeval star formation to within 2 -- 3 Myr for both LH~51 and LH~54.
The IMF agrees with the Salpeter (1955) value within the large
uncertainties, for all the subgroups.

We have reexamined the inferred input wind power $L_w(t)$, by computing
$L_w(t)$ from stellar $\Mdot$ relations of Puls {\etal}(1996) and Howarth
\& Prinja (1989).  The $\Mdot$ and $L_w(t)$ from these relations differ
substantially from those of de Jager {\etal}(1988), whose relations
are used in the Geneva stellar evolution models.  While the JNH88
models increase by an order of magnitude over the H-burning phase,
those of P+96 and HP89 remain almost constant, yielding a slightly
different shell evolution.  However, it is necessary to use stellar
models that incorporate these $\Mdot$ to reliably obtain $L_w(t)$
from these models, since $\Mdot$ could significantly affect
the stellar evolution timescales.

We also investigated the effect of a sudden drop in ambient density on
the modeled shell evolution.  Thus far, all the objects studied in
this series, including DEM~192, show enhanced X-ray emission and a
depleted IMF, suggesting SN acceleration of the shells.  However, a
discontinuous drop in $n$ by a factor of a 
few at a given distance from the origin can fully reproduce the shell
kinematics seen in the ``high-velocity'' superbubbles such as DEM 192
and objects studied by Oey (1996a).  Shell acceleration by SNR impacts
are therefore not necessary to explain a high observed $v_s$ in
comparison to $R_s$.  


\acknowledgments
We thank Becky Elson for useful discussions and Joachim Puls for
providing the calibration of the wind momentum -- stellar luminosity
relation.  We are also grateful to Marcus Bruggen for software
assistance.  Thanks to the anonymous referee for comments and suggestions.
Some of this work was carried out by MSO at 
the University of Arizona, where she received support from NSF grants 
AST90-19150 and AST94-21145, the American Association of University
Women, and the University of Arizona.

\end{document}